\documentclass[11pt]{article}

\usepackage[a4paper,top=2cm,bottom=2cm,left=2.5cm,right=2.5cm,marginparwidth=1.75cm]{geometry}

\usepackage{amsfonts,amsmath,amssymb,ascmac,bm,tensor}
\usepackage{acronym, cite}

\usepackage{graphicx, color}     %   usepackage without driver option
\ifpdf
  \usepackage[bookmarksopen,colorlinks=true,linkcolor=blue,citecolor=blue,urlcolor=magenta]{hyperref}
\else     % For (p)LaTeX + dvipdfmx
\fi

\makeatletter
\let\MYcaption\@makecaption
\makeatother

\usepackage{subcaption}
\makeatletter
\let\@makecaption\MYcaption
\makeatother

\allowdisplaybreaks

\title{F-type multi-field inflation in supergravity without stabiliser superfields}

\begin{document}

\begin{titlepage}
\begin{center}

December 2025  \hfill{APCTP Pre2025-017}\\
\hfill{IPMU25-0040}

\vspace{2.6 cm}

{\LARGE\bf $F$-term Multi-Field Inflation in Supergravity \\ without Stabiliser Superfields}

\vspace{1.4 cm}

{\large Jinn-Ouk Gong$^{a, b}$, Sergei V. Ketov$^{c, d, e}$, and Takahiro Terada$^{f}$}

\vspace{7 mm}

${}^{a}$ Department of Science Education, Ewha Womans University, Seoul 03760, Korea \\
${}^{b}$ Asia Pacific Center for Theoretical Physics, Pohang 37673, Korea \\
${}^{c}$ Department of Physics, Tokyo Metropolitan University, Tokyo 192-0397, Japan \\
${}^{d}$ Kavli Institute for the Physics and Mathematics of the Universe (WPI),\\
The University of Tokyo Institutes for Advanced Study, Chiba 277-8583, Japan \\
${}^{e}$ Interdisciplinary Research Laboratory, Tomsk State University, Tomsk 634050, Russia \\
${}^{f}$ Kobayashi-Maskawa Institute for the Origin of Particles and the Universe, \\
Nagoya University, 
Nagoya 464-8602, Japan %\\

\vspace{1cm}

\normalsize 

\begin{abstract}

\noindent Realising $F$-term slow-roll inflation in supergravity is non-trivial due to the well-known $\eta$-problem. 
The common strategy to solve the problem is to impose a shift symmetry on the K\"ahler potential,
 but this often leads to a negative potential in the large-field regime. To avoid negative potentials, an additional 
 superfield called the stabiliser is usually added with a desired interaction.  
 An alternative mechanism in supergravity, avoiding the use of a stabiliser superfield, was earlier proposed by two of us in the setup with a single chiral superfield having inflaton and goldstino amongst its field components. In this work, we extend that alternative mechanism to multi-superfield  models of inflation, thereby providing a generic framework for embedding a wide class of single- and multi-field inflation models into supergravity. We illustrate our approach by several concrete examples of multi-field inflation and clarify the conditions required to avoid tachyonic instabilities during multi-field evolution. Our proposal significantly broadens the theoretical  landscape of $F$-term inflation models in supergravity.
\end{abstract}

\end{center}

\end{titlepage}

%======================================================================
\section{Introduction}
\label{sec:intro}

Despite overwhelming (indirect) evidence for cosmological inflation in the early Universe via precision observations of the cosmic microwave background (CMB) radiation and strong theoretical support from resolving fundamental problems of the standard (Einstein-Friedmann) cosmology, a detailed microscopic mechanism of inflation is still unknown. 

The simplest theoretical models of inflation use a real scalar field called the inflaton, whose action is given by 
quintessence with a particular scalar potential or, equivalently, a modified $F(R)$ gravity theory.  On one hand, since CMB is a small (in scales) window to inflation, there exist many viable models of inflation in the literature~\cite{Martin:2024qnn}. On the other hand, all models of inflation are becoming more constrained by CMB measurements by Planck/BICEP/Keck  Collaborations~\cite{Planck:2018vyg, Planck:2018jri, BICEP:2021xfz}, Atacama Cosmology Telescope (ACT)~\cite{ACT:2025fju,ACT:2025tim}, and South Pole Telescope (SPT-3G)~\cite{SPT-3G:2025bzu}, combined with large-scale structure observations such as Dark Energy Spectroscopic Instrument (DESI)~\cite{DESI:2024mwx, DESI:2024hhd, DESI:2025zgx}. 
Since inflation is sensitive to quantum gravity, there are additional constraints on the single-field realisations of inflation coming from the Swampland Program \cite{Palti:2019pca}, see \textit{e.g.}, Refs.~\cite{Scalisi:2018eaz,Ketov:2025nkr}. The constraints on single-field models of inflation become even more severe after adding a production of primordial black holes as dark matter \cite{Carr:2020gox, Carr:2020xqk, Green:2020jor, Escriva:2022duf} on smaller (than CMB) scales \cite{Frolovsky:2025iao}.

However, there is no compelling reason to rely on single-field models of inflation besides their simplicity, while there is no
good reason for the existence of only one fundamental (Higgs) scalar field up to the inflationary energy scale as well. Though the latest CMB data do not show any compelling signals of isocurvature perturbations and non-Gaussianity,  and single-field models are consistent with that,  this does not mean multi-field inflation is excluded, because there are many multi-field models without significant isocurvature perturbations and non-Gaussianity. Therefore, multi-field inflation is a viable alternative to single-field inflation, see \textit{e.g.}, Refs.~\cite{Langlois:2008mn,Gong:2016qmq,Ketov:2021fww}. Though 
multi-field inflation is well supported by theoretical physics beyond the Standard Model of elementary particles, and by
string theory as quantum gravity as well, it has much less predictive power in general because of the proliferation of fields and couplings. Therefore, some guidance is needed to restrict multi-field inflation by fundamental theoretical principles in addition
to CMB observations that are easy to match in the presence of many free parameters. 

Supersymmetry is a fundamental theoretical principle that can restrict high-scale multi-field inflation without a conflict with
Large Hadron Collider measurements on electroweak scales. In the context of inflation, supersymmetry should be locally
realised because  then it automatically implies general coordinate invariance, which leads to supergravity. The $\mathcal{N}=1$
supergravity in curved four-dimensional spacetime is known as the natural framework for the unification of fundamental forces
with gravity, high-scale inflation, primordial black holes production, reheating, leptogenesis and baryogenesis \cite{Ellis:2020lnc,Jeong:2023zrv,Ketov:2023ykf}. 

In supersymmetry, the inflaton scalar should be assigned to an irreducible massive $\mathcal{N}=1$ supermultiplet that can be either a chiral one or a vector one. If the inflaton multiplet is chiral (\textit{i.e.} described by a chiral superfield),  its physical scalar field component must be complex,  \textit{i.e.} real inflaton must be accompanied by the pseudo-scalar superpartner called sinflaton (or vice versa). 
The scalar potential is given by the so-called $F$-term, where $F$ is the auxiliary scalar of the chiral superfield and an order parameter of supersymmetry breaking. We employ the $F$-term scalar potential in this paper, without using gauge 
fields.\footnote{A massive vector supermultiplet has a real massive physical scalar that is suitable as the inflaton for embedding some single-field inflation models whose scalar potential is given by a real function squared, known as the $D$-term, where $D$ is the auxiliary field of the vector supermultiplet and the order parameter of supersymmetry breaking as well~\cite{Farakos:2013cqa, Ferrara:2013rsa, Ketov:2019toi}. Multi-field inflation with multiple or non-Abelian vector supermultiplets is beyond the scope of this work.} 

Inflation in supergravity has another generic feature related to a positive energy driving inflation, which implies spontaneous
supersymmetry breaking during inflation. In turn, spontaneous supersymmetry breaking implies the existence of the Nambu-Goldstone-type fermionic (or spin 1/2) field known as goldstino, which is absorbed by spin 3/2 gravitino by the super Higgs mechanism.  In general, goldstino is a  fermionic field component of a chiral superfield that can be  different from the chiral inflaton superfield. In such cases~\cite{Kawasaki:2000yn, Kawasaki:2000ws, Kallosh:2010ug, Kallosh:2010xz}, the goldstino superfield is often called a stabiliser superfield in the literature on inflation in supergravity. 

It is possible to identify the inflaton superfield with the goldstino superfield so that both inflaton and goldstino become superpartners, which is a more economical setup \cite{Ketov:2014qha} (see Refs.~\cite{Izawa:2007qa, Achucarro:2012hg, Alvarez-Gaume:2010fpq, Alvarez-Gaume:2011gxt} for earlier attempts). 
Also, in Ref.~\cite{Ketov:2014qha}, the mechanism for phenomenological realisation of inflation in supergravity with a single chiral superfield was proposed without using a 
stabiliser superfield. The proposal in Ref.~\cite{Ketov:2014qha} was not a specific inflation model but the mechanism that can accommodate many inflation models in supergravity. The mechanism was further developed in Refs.~\cite{Ketov:2014hya, Terada:2014gnl, Ketov:2015tpa, Ketov:2016gej} and then related physical aspects were also studied, \textit{e.g.}, in Refs.~\cite{Terada:2014gnl, Linde:2014ela, Ema:2016oxl, Ema:2018jgc, Gao:2018pvq, Aoki:2021nna}.  In this paper, we extend the setup of Ref.~\cite{Ketov:2014qha} to multi-superfield generalisations of inflation without a stabiliser field, where goldstino is a linear combination of the fermionic field components of several chiral superfields.

The paper is organised as follows. In Sec.~\ref{sec:review}, the well-known basic problems on the realisation of inflation in supergravity are briefly reviewed. In Sec.~\ref{sec:generalization}, we propose the multi-field model building in supergravity without stabiliser fields, demonstrate how to solve the basic problems, and provide some specific examples. Sec.~\ref{sec:stabilization} is devoted to a stabilisation mechanism for 
multi-superfield inflation. Our conclusion is in Sec.~\ref{sec:conclusion}. 
Our notation involving field-space basis rotation is clarified in Appendix~\ref{sec:field_space_rotation}. Since we briefly discuss $R$-axion in Sec.~\ref{sec:generalization}, the constraints on $R$-axions are summarised in Appendix~\ref{sec:R-axion_constraints}. 

Throughout the paper, we use the natural units $c=\hbar=1$, as well as the reduced Planck mass 
$M_\text{Pl}^2 = 1/(8\pi G)=1$ unless the latter is explicitly written.  

%======================================================================
\section{Inflation in supergravity}
\label{sec:review}

In this Section, we briefly address three problems in supergravity, namely, (i) how to get a positive potential during inflation, (ii) how to realise a shift symmetry needed for approximate flatness of the potential during inflation, and (iii) how
to take into account the higher-order terms. See Refs.~\cite{Yamaguchi:2011kg, Terada:2014gnl} for more details.

\subsection{Bosonic action in supergravity}

In the absence of gauge fields, supergravity models are defined by a non-holomorphic, Hermitian K\"{a}hler potential $K=K(\Phi^i, \bar{\Phi}^{\bar{i}})$ and a holomorphic superpotential $W=W(\Phi^i)$, where $\Phi^i$ denote chiral superfields as well as their leading bosonic field components, and $\bar{\Phi}^{\bar{i}}$ denote their Hermitian conjugates and their leading bosonic field components as well. The fermionic contributions are usually ignored in supergravity models of inflation because 
possible fermionic contributions to correlators of perturbations are highly suppressed during inflation.

The bosonic part of the supergravity Lagrangian (in the Einstein frame) after eliminating the auxiliary fields reads 
\begin{align}
    \mathcal{L} = \sqrt{-g} \left[ \frac{1}{2}R  - K_{i\bar{j}} g^{\mu\nu}\partial_\mu \bar{\Phi}^{\bar{j}} \partial_\nu \Phi^i  
    - V(\Phi^i, \bar{\Phi}^{\bar{j}})\right] ,
\end{align}
where $R$ is the spacetime Ricci scalar curvature and $K_{i\bar{j}} \equiv \partial^2 K / \partial \Phi^i \partial \bar{\Phi}^{\bar{j}}$ is the K\"ahler metric in field space of the non-linear sigma model  \cite{Ketov:2000dy}.
The corresponding scalar potential is given by
\begin{align}\label{pot}
    V = e^{K} \left[ K^{\bar{j}i} (W_i + K_i W) (\overline{W}_{\bar{j}} + K_{\bar{j}} \overline{W}) - 3 |W|^2 \right],
\end{align}
where $K^{\bar{j}i}$ is the inverse of the K\"ahler metric, the subscripts denote the derivatives with respect to the corresponding fields, as  \textit{e.g.}, $W_i \equiv \partial W/\partial \Phi^i$, and the bars denote Hermitian conjugation.  

To realise slow-roll inflation, one needs a positive and sufficiently flat potential. It is non-trivial to meet both conditions with
Eq.~(\ref{pot}). The first obstacle to inflation in supergravity is the exponential factor $e^K$. For example, the minimal K\"ahler potential $K = |\Phi^i|{}^2$ leads to an exponential potential $V\sim e^{|\Phi^i|^2}$ that  is too steep for slow-roll inflation. This is known as  the $\eta$-problem in supergravity inflation~\cite{Copeland:1994vg}. In small-field inflation, fine-tuning of parameters at a point in the field space may be enough to solve the $\eta$-problem, but for large-field inflation, fine-tuning along the whole inflationary trajectory is needed. 
Another obstacle is the presence of the negative semi-definite term $-3e^K |W|^2$ in Eq.~(\ref{pot}). This term may not look like a severe problem at first glance, but it becomes problematic after solving the first obstacle by introducing a shift symmetry, see
below.

\subsection{Shift symmetry and stabiliser superfield}

One way to solve the steepness issue is to introduce an approximate shift symmetry under the shift transformations $\Phi \to \Phi + a$.  Let us suppose the real part of $\Phi$ corresponds to the inflaton.  In this case, the constant $a$ should be the real transformation parameter.  When imposing the invariance of a K\"ahler potential under this transformation, we get
\begin{align}
  K(\Phi, \bar{\Phi}) = K(\Phi - \bar{\Phi}),
\end{align}
where we have focused on the dependence upon the inflaton superfield $\Phi$. Then $K$ depends only on the imaginary part of 
$\Phi$ or sinflaton, so that the factor $e^K$ does not depend on the real part, which is the inflaton.

However, this also implies that $K_i$ depends only on the imaginary part that vanishes or is suppressed in many simple models of inflation. Then the large-field behaviour of the potential looks like $V \sim K^{\bar{j}i}W_i \bar{W}_{\bar{j}} - 3 |W|^2$, where for generic (\textit{e.g.}, polynomial) $W(\Phi)$ the second term will dominate.\footnote{
This can be evaded by exponential superpotentials. To obtain a sufficiently flat potential, however, a delicate balance between multiple exponential terms is required~\cite{Ellis:1983ei, Goncharov:1983mw, Roest:2015qya, Linde:2015uga}.  See also Refs.~\cite{Ferrara:2016vzg, Ellis:2018xdr, Ellis:2019hps} for more recent developments.
} As a result, the potential becomes negative in the large-field region, namely, $V \sim - 3 |W|^2 < 0$.

To solve this problem, another superfield was introduced in Refs.~\cite{Kawasaki:2000yn, Kawasaki:2000ws}, which is often called the stabiliser superfield in the literature. 
The idea is to introduce a field $S$ that vanishes during inflation and assume $W\propto S$.  This implies $W$ vanishes during inflation on the equations of motion.  The inflation energy comes from the 
$|W_S|^2$ term that leads to the $F$-term supersymmetry breaking.

More specifically, the corresponding class of inflation models is defined by~\cite{Kawasaki:2000yn, Kawasaki:2000ws, Kallosh:2010ug, Kallosh:2010xz}
\begin{align}
    K &= -\frac{1}{2} (\Phi - \bar{\Phi})^2 + |S|^2 + \cdots, \\
    W &= S f(\Phi),
\end{align}
where the dots stand for additional higher-order terms needed to stabilise $S$ (usually near $S=0$),  leading to a more general K\"ahler potential, and $f(\Phi)$ is an arbitrary holomorphic function.  Having the stabilised $S = 0$ and neglecting the dots in the $K$ above, one arrives at the positive scalar potential 
\begin{align}
V = e^{K} |f(\Phi)|^2.
\end{align}

\subsection{Shift-symmetric higher-order terms}

Introducing a stabiliser superfield is not the only solution. It is possible to avoid additional physical degrees of freedom for large-field inflation in supergravity~\cite{Ketov:2014qha} by exploiting the shift symmetry again and boosting the size of either $K_i$ or $W_i / W$. 

Let us consider a generic $W$ with the following shift-symmetric $K$:
\begin{align}
K(\Phi, \bar{\Phi})  = i c (\Phi - \bar{\Phi}) - \frac{1}{2}(\Phi - \bar{\Phi})^2  - \frac{\xi}{12}(\Phi - \bar{\Phi})^4 + \cdots, 
\end{align}
where $c$ and $\xi$ are real parameters. The quadratic term gives the approximately canonical kinetic term.  The quartic term ensures the imaginary part of $\Phi$ approximately vanishes as long as $\Phi$ breaks supersymmetry and $\xi > +\mathcal{O}(1)$ because it leads to supersymmetry-breaking mass squared term $12 \xi (H^2 + m_{3/2}^2)$, where $H$ is the Hubble parameter and $m_{3/2}=e^{K/2}|W|$ is the gravitino mass. 
Then, the first term leads to a non-vanishing $K_\Phi$, which boosts the positive contribution to the scalar potential. 
The absence of the cubic term can be understood as follows. Suppose there is initially a cubic term as well as other shift-symmetric terms. Provided that the quartic term is sufficiently large, the imaginary part is stabilised at a finite field value. Expanding $K$ around this point, the cubic term approximately vanishes. Because of the shift symmetry in $K$, the flat direction $\text{Re}\, \Phi$ is the inflaton candidate, whose potential is specified by the superpotential $W$, which explicitly breaks the shift symmetry.   

The corresponding scalar potential is given by
\begin{align}
    V = e^K \left\{ K^{\Phi \bar{\Phi}} \Big[ |W_\Phi|^2 + i c \big(W \overline{W}_{\bar{\Phi}} - \overline{W} W_\Phi \big) + c^2 |W|^2 \Big] - 3 |W|^2 \right\}.
\end{align}
To avoid the negative term from dominating, we impose the condition
\begin{align}
    K^{\Phi\bar{\Phi}} c^2 \geq 3. \label{condition_c_single}
\end{align}
In the above parametrisation we have $K^{\bar{\Phi}\Phi} \approx 1$, so that $|c| \gtrsim \sqrt{3}$. 

The equivalent idea is to enhance the derivative of the superpotential relative to the superpotential itself. For superpotentials as generic polynomials, this is not possible in the large-field region, but it becomes possible via adding an exponential factor,  
\begin{equation}\label{single}
    K(\Phi, \bar{\Phi}) = - \frac{1}{2}(\Phi - \bar{\Phi})^2  - \frac{\xi}{12}(\Phi - \bar{\Phi})^4 + \cdots, \qquad
    W(\Phi) = e^{ic \Phi} \widetilde{W}(\Phi),
\end{equation}
where  $\widetilde{W}$ is another holomorphic function.

Compared to the previous $K$ and $W$, no linear term appears in the K\"ahler potential, but there is the exponential factor in the superpotential instead. Both descriptions are equivalent because of the super-Weyl-K\"ahler invariance in supergravity: Two theories specified by $(K,W)$ and $(K + F + \bar{F},e^{-F}W)$ are equivalent up to quantum anomalies~\cite{Bagger:1999rd}, where $F$ is a holomorphic function. 
The condition for $c$ to avoid the negative potential in the large-field region is the same as Eq.~\eqref{condition_c_single}.  The exponential factor makes the imaginary direction exponentially steep, but it does not introduce the exponential dependence in the real (inflaton) direction because it is just an overall phase factor. 

These apparently different descriptions can be unified as long as $W$ does not vanish in terms of the ``total'' K\"ahler potential $G \equiv K + \log |W|^2$, which is invariant under the super-Weyl-K\"ahler transformation. The above setup requires that $G$ contains a shift-symmetric linear term $i c (\Phi - \bar{\Phi})$ and the appropriate K\"ahler curvature, which is essentially the negative quartic term in $G$ (or equivalently in $K$).  

Various extensions and applications of the above mechanism were studied in the literature.  The special form of the K\"ahler potential was proposed in Ref.~\cite{Ketov:2014hya} in order to embed an arbitrary non-negative scalar potential to supergravity. The alternative description in terms of a $U(1)$ symmetry instead of the shift symmetry was proposed in Ref.~\cite{Ketov:2015tpa}. The generic way of restoring supersymmetry after inflation was given in Ref.~\cite{Ketov:2016gej}.  
Gravitino production in supergravity models with and without a stabiliser field was studied in Ref.~\cite{Ema:2016oxl}. The relation between the sign of the quartic term in the K\"ahler potential and unitarity for some ultraviolet completion was studied in Ref.~\cite{Ema:2018jgc}.  Primordial black hole production in this context was discussed in Ref.~\cite{Gao:2018pvq}.  The low-energy effective field theory of inflation with a single superfield, where supersymmetry is non-linearly realised, was identified in Ref.~\cite{Aoki:2021nna}.

%======================================================================
\section{Multi-field generalisation of supergravity inflation}
\label{sec:generalization}

Inflation in the models of the preceding Section was essentially driven by a single inflaton field.
Let us now generalise those models to several superfields in the inflaton sector  by introducing
an extension inspired by Eq.~(\ref{single}) as follows:
\begin{equation}\label{many}
    K(\Phi^i, \bar{\Phi}^{\bar i}) =  K(\Phi^i - \bar{\Phi}^{\bar{i}}), \qquad
    W(\Phi^i) =  e^{i c_j \Phi^j} \widetilde{W}(\Phi^i) \, ,
\end{equation}
where the shift symmetries were imposed on every chiral superfield $\Phi^i$. One can imagine that fields not protected by the shift symmetry become heavy and decoupled. In this sense, the real parts of $\Phi^i$ are light during inflation, and they become candidates of the inflaton. We assume that there are sufficiently large negative quartic terms in $K$ so that the imaginary parts are stabilised. This point is further studied and justified in Sec.~\ref{sec:stabilization}. The overall exponential dependence $\exp (i \sum_j c_j \Phi^j)$ was extracted from the superpotential, leaving a holomorphic function $\widetilde{W}$. We assume all $c_j$'s to be real, $c_j = c_{\bar{j}}$.  For a later use, we note $W_i = (i c_i \widetilde{W} + \widetilde{W}_i) e^{i c_j \Phi^j}$.

The above generalisation is analogous to the second approach in the preceding Section. Equivalently, it is possible to replace the exponential factor with a linear term in $K$ by the K\"ahler transformation, $K \to K + \Delta K$, with
\begin{align}
    \Delta K = i c_j (\Phi^j - \bar{\Phi}^{\bar{j}}),
\end{align}
and the superpotential becomes $\widetilde{W}$ instead of $W=e^{ic_j \Phi^j} \widetilde{W}$. 
Similar to the single-superfield case, the super-Weyl-K\"ahler invariant description can be given in terms of the total K\"ahler potential $G$, and the crucial ingredients are the linear and quartic shift-symmetric terms in $G$.

The scalar potential reads
\begin{align}
V = e^{K}  e^{i c_k (\Phi^k - \bar{\Phi}^{\bar{k}} )}  \left[ K^{\bar{j}i} \left( \widetilde{W}_i \overline{\widetilde{W}}_{\bar{j}} + i \tilde{c}_i \widetilde{W}\overline{\widetilde{W}}_{\bar{j}} - i \tilde{c}_{\bar{j}} \overline{\widetilde{W}}\widetilde{W}_i + \tilde{c}_i \tilde{c}_{\bar{j}} |\widetilde{W}|^2 \right) - 3 |\widetilde{W}|^2 \right],
\end{align}
where we have defined $i \tilde{c}_i \equiv i c_i + K_i$ and $-i\tilde{c}_{\bar{j}} \equiv 
 - i c_{\bar{j}} + K_{\bar{j}}$. 
Note that the factors like $e^K$, $e^{i c_k (\Phi^k - \bar{\Phi}^{\bar{k}} )}$ and $K^{\bar{j}i}$ respect the shift symmetries.
When all the coupling constants in $W$ are real, the interference terms between the zeroth and first derivatives of $W$ cancel,
and only the combination $\widetilde{W}_i \overline{\widetilde{W}}_{\bar{j}} + \tilde{c}_i \tilde{c}_{\bar{j}} |\widetilde{W}|^2$ is multiplied by $K^{\bar{j}i}$.

For a generic (\textit{e.g.}, polynomial) functions $\widetilde{W}(\Phi^i)$, the terms proportional to $|\widetilde{W}|^2$ grow faster than the others in the large-field region. 
Hence, demanding the coefficient at $|\widetilde{W}|^2$ be non-negative imposes the inequality condition
\begin{align}
    K^{\bar{j}i} \tilde{c}_i \tilde{c}_{\bar{j}} \geq 3.
\end{align}
This is a generalisation of Eq.~\eqref{condition_c_single}.

In the rest of this Section, we provide some example choices of superpotential to demonstrate that this multi-field generalisation accommodates various inflation models.  For simplicity, we work in the ideal stabilisation limit ($\xi \gg 1$), in which the imaginary parts vanish (see Sec.~\ref{sec:stabilization} for the stabilisation mechanism). Our focus in this paper is on the theoretical formulation of embedding generic inflation models into supergravity, so we leave detailed observational implications of the following example models to future work, such as estimation of the spectral index $n_\text{s}$, tensor-to-scalar ratio $r$, isocurvature perturbations, and non-Gaussianity, which will require extensive numerical calculations of multi-field dynamics.  For some examples below, we can see without detailed calculations that they are incompatible with the latest observations. But again, this is because our purpose is to illustrate that the present mechanism is flexible enough to realise representative inflationary models.

\subsection{Nearly constant superpotential}

It is instructive to consider the simple case of constant $\widetilde{W}$ as a prototype for de Sitter-like evolution with
\begin{align}
    \widetilde{W} = \widetilde{W}_0 = \text{constant}.
\end{align}
The scalar potential becomes
\begin{align}
    V = e^K e^{i c_k (\Phi^k - \bar{\Phi}^{\bar{k}})} \left( K^{\bar{j}i}\tilde{c}_i \tilde{c}_{\bar{j}} - 3\right) |\widetilde{W}_0|^2.
\end{align}
The overall factors do not depend on the real parts of the scalar fields, so they reduce to some constant after stabilisation of the imaginary parts.  We can assume that the combined factor $e^K e^{i c_k (\Phi^k - \bar{\Phi}^{\bar{k}})}$ to be $1$ without loss of generality by redefining $\widetilde{W}_0$. We also assume during inflation that the other factors like $\tilde{c}_i$ and $K^{\bar{j}i}$ are also (at least approximately) constant. Then, provided that $K^{\bar{j}i}\tilde{c}_i \tilde{c}_{\bar{j}} > 3$, de Sitter spacetime is realised due to a positive cosmological constant. 

This motivates us to modify $\widetilde{W}$ to obtain quasi-de Sitter, slow-roll inflation. With two superfields $\Phi^1 = \Phi$ and $\Phi^2 = \Psi$, we deform $\widetilde{W}$ as follows:
\begin{align}
    \widetilde{W} = \widetilde{W}_0\left( 1 - a_\Phi e^{-b_\Phi \Phi} - a_\Psi e^{-b_\Psi \Psi} - a_{\Phi\Psi} e^{- b'_\Phi \Phi - b'_\Psi \Psi}\right), \label{W_plateau}
\end{align}
where the parameters $b_\Phi$, $b_\Psi$, $b'_\Phi$, and $b'_\Psi$ are real and positive while the other parameters $a_\Phi$, $a_\Psi$, and $a_{\Phi\Psi}$ will be specified right below. Note that the superpotential asymptotes to a constant as we discussed above in the limit $\Phi, \Psi \to +\infty$, where de Sitter spacetime can be realised.   
We set $\Phi = \Psi = 0$ at the minimum of the potential without loss of generality.

Next, we impose the conditions $W = W_\Phi = W_\Psi = 0 $ at $\Phi = \Psi = 0$ after inflation, which is a sufficient condition to restore supersymmetry after inflation (at least at the inflationary scale)~\cite{Ketov:2016gej}. The explicit form of these conditions is as follows:
\begin{align}
a_\Phi + a_\Psi + a_{\Phi \Psi} =& 1, \\
a_\Phi b_\Phi + a_{\Phi \Psi} b'_\Phi = & 0, \\
a_\Psi b_\Psi + a_{\Phi \Psi} b'_\Psi = & 0.
\end{align}
With the given set of exponents, we fix $a_\Phi$, $a_\Psi$, and $a_{\Phi \Psi}$ from these conditions as follows:
\begin{align}
    a_\Phi =  &\frac{b'_\Phi b_\Psi}{b_\Phi b'_\Psi + b'_\Phi b_\Psi - b_\Phi b_\Psi}, \\
    a_\Psi =  &\frac{b_\Phi b'_\Psi}{b_\Phi b'_\Psi + b'_\Phi b_\Psi - b_\Phi b_\Psi}, \\
    a_{\Phi \Psi} = &\frac{- b_\Phi b_\Psi}{b_\Phi b'_\Psi + b'_\Phi b_\Psi - b_\Phi b_\Psi}. 
\end{align}

\begin{figure}[tbhp]
    \centering
    \includegraphics[width=0.6\linewidth]{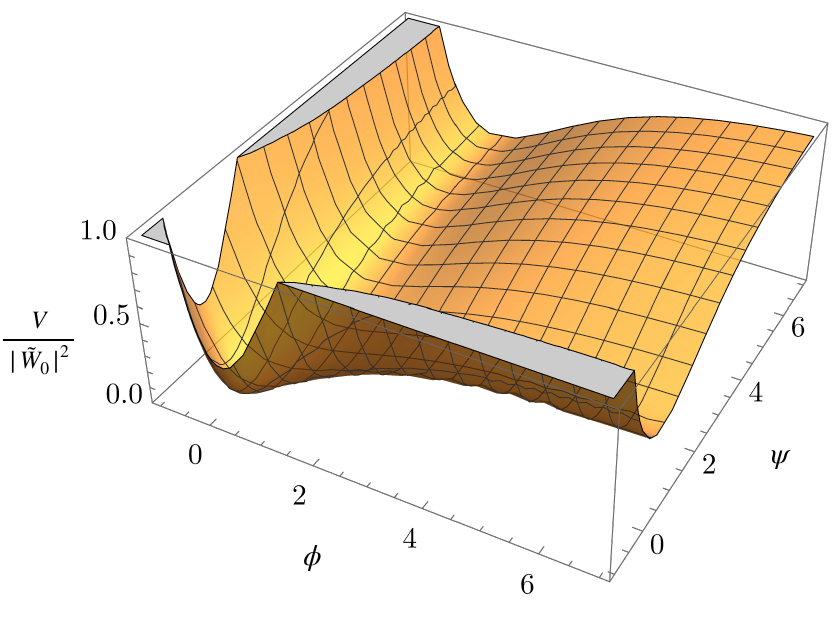}
    \caption{An example of the normalized scalar potential~\eqref{V_multi_exp} in terms of $\phi = \sqrt{2} \text{Re}\,\Phi$ and $\psi = \sqrt{2} \text{Re}\,\Psi$. The parameters are taken as $b_\Phi = b_\Psi = b'_\Phi = b'_\Psi = \sqrt{2/3}$ and $c = 2$. 
    }
    \label{fig:V_multi_exp}
\end{figure}

The scalar potential after stabilisation reads
\begin{align}
    \frac{V}{|\widetilde{W}_0|^2}=& \left|a_\Phi b_\Phi e^{- b_\Phi \Phi} + a_{\Phi \Psi} b'_\Phi e^{-b'_\Phi \Phi - b'_\Psi \Psi}\right|^2  + \left|a_\Psi b_\Psi e^{- b_\Psi \Psi}+ a_{\Phi\Psi} b'_\Psi e^{-b'_\Phi \Phi - b'_\Psi \Psi}\right|^2 \nonumber \\
    & + (c^2 -3) \left|1 - a_\Phi e^{-b_\Phi \Phi} - a_\Psi e^{-b_\Psi \Psi} - a_{\Phi \Psi} e^{-b'_\Phi \Phi - b'_\Psi \Psi}\right|^2, \label{V_multi_exp_complex}
\end{align}
where $c^2 \equiv c^2_\Phi + c^2_\Psi$. 
In this example, the parameters $a$'s, $b$'s, and $c$ are all real whereas the imaginary parts are stabilised, so the potential is further simplified in terms of two real scalar fields $\phi = \sqrt{2} \text{Re}\,\Phi$ and $\psi = \sqrt{2}\text{Re}\, \Psi$ as follows:
\begin{align}
    \frac{V}{|\widetilde{W}_0|^2}=& \left(a_\Phi b_\Phi e^{- b_\Phi \phi/\sqrt{2}} + a_{\Phi \Psi} b'_\Phi e^{-(b'_\Phi \phi + b'_\Psi \psi)/\sqrt{2}}\right)^2  + \left(a_\Psi b_\Psi e^{- b_\Psi \psi/\sqrt{2}}+ a_{\Phi\Psi} b'_\Psi e^{-(b'_\Phi \phi + b'_\Psi \psi)/\sqrt{2}}\right)^2 \nonumber \\
    & + (c^2 -3) \left(1 - a_\Phi e^{-b_\Phi \phi/\sqrt{2}} - a_\Psi e^{-b_\Psi \psi/\sqrt{2}} - a_{\Phi \Psi} e^{-(b'_\Phi \phi + b'_\Psi \psi)/\sqrt{2}}\right)^2. \label{V_multi_exp}
\end{align}
A profile of this potential is plotted in Fig.~\ref{fig:V_multi_exp} that qualitatively describes two-field inflation.

\subsection{Hybrid inflation}

Hybrid inflation~\cite{Linde:1993cn} is one of the standard types of multi-field models of inflation in the literature.  
To realise hybrid inflation in our supergravity setup, let us  consider the following superpotential:
\begin{align}
    \widetilde{W}= \lambda \Phi (M^2 - \Psi_1 \Psi_2)
\end{align}
in terms of three chiral superfields $\Phi$, $\Psi_1$, and $\Psi_2$ with the parameters $M$ and $\lambda$. 

The last term could have been replaced by $\Psi^2$, but we find it more useful to use the oppositely charged scalar fields $\Psi_1$ and $\Psi_2$ from the model-building viewpoint.  Here, $\Phi$ plays the role of the inflaton while $\Psi_1$ and $\Psi_2$ play the role of the waterfall fields. The overall phase of superpotential is unphysical, 
so we take $\lambda$  to be real.\footnote{Extra care is needed here because a phase rotation of $\Psi_1$ and $\Phi_2$ may be incompatible with the shift symmetry structure $(\Psi_{1,2} - \bar{\Psi}_{1,2})$ in $K$, so we take $M^2$ as complex.}

The scalar potential takes the form
\begin{align} \label{pot3}
    \frac{V}{\lambda^2} =& |M^2 - \Psi_1 \Psi_2|^2 + |\Phi|^2 (|\Psi_1|^2 + |\Psi_2|^2) + (c^2 - 3) |\Phi|^2 |M^2 - \Psi_1 \Psi_2|^2,
\end{align}
where $c^2 \equiv c_\Phi^2 + c_{\Psi_1}^2 + c_{\Psi_2}^2$.
The first term in Eq.~(\ref{pot3}) is responsible for the vacuum structure and the tachyonic masses of the waterfall fields. The second term is the inflaton-dependent contribution to the masses of the waterfall fields, which can be used to stabilise them when $\Phi$ is sufficiently large.  The third term gives the inflaton-dependent corrections to the scalar potential and to the masses of the waterfall fields. The quadratic terms with respect to the waterfall fields are
\begin{align}
    \frac{V}{\lambda^2} = \begin{pmatrix}
        \bar{\Psi}_1 & \Psi_2
    \end{pmatrix}
    \begin{pmatrix}
        |\Phi|^2 & -M^2 \left( 1 + (c^2 - 3)|\Phi|^2 \right) \\
        -\bar{M}^2 \left( 1 + (c^2 - 3)|\Phi|^2 \right) & |\Phi|^2 
    \end{pmatrix}
    \begin{pmatrix}
        \Psi_1 \\ \bar{\Psi}_2
    \end{pmatrix}+ \cdots.
\end{align}
The determinant of the mass matrix becomes negative with a negative eigenvalue for $|\Phi| < |\Phi_\text{c}|$ with 
\begin{align}
    |\Phi_\text{c}| =& \frac{|M|}{\sqrt{1 - (c^2 -3) |M|^2}}~.
\end{align}
Along the inflation trajectory with $\Psi_1 = \Psi_2 = 0$ we find
\begin{align}\label{pot4}
    \frac{V}{\lambda^2 |M|^4} = 1  + (c^2 - 3) |\Phi|^2.
\end{align}
Therefore, for the inflaton to roll down toward the waterfall critical point, the condition $c^2 - 3 \geq0$ should be satisfied,
as should be expected from the preceding Section.

The potential \eqref{pot4} does not obey the CMB observational constraints on the power spectrum of perturbations (on the tree level) due to its simple quadratic form. 
However, it can be easily generalised with more parameters, such as
\begin{align}
    \widetilde{W} = \mu (1 - a e^{-b \Phi}) (M^2 - \Psi_1 \Psi_2),
\end{align}
where $\mu$ and $b$ are taken as real and positive.  
Then, the scalar potential is
\begin{align}
    \frac{V}{\mu^2} & = |a b|^2 e^{- b (\Phi + \bar{\Phi})} |M^2 - \Psi_1 \Psi_2|^2 + | 1- a e^{-b \Phi}|^2 (|\Psi_1|^2 + |\Psi_2|^2)  \nonumber \\
    & \quad + (c^2 - 3) | 1- a e^{-b \Phi}|^2 |M^2 - \Psi_1 \Psi_2|^2. 
\end{align}
Similar to the previous example, the system is described by three real scalar fields $\phi = \sqrt{2}\text{Re}\, \Phi$ and $\psi_{1,2} = \sqrt{2}\text{Re}\, \Psi_{1,2}$. 
In the large-field limit $b \phi \gg1$, 
$V/\mu^2 \approx |\Psi_1|^2 + |\Psi_2|^2 + (c^2 -3)|M^2 - \Psi_1 \Psi_2|^2$, the first two terms can be dominant over the last Planck-suppressed term. Hence, we can set $\Psi_1 = \Psi_2 = 0$ during inflation. Then the inflaton potential becomes $V /(\mu^2 |M|^4) = |a b|^2 e^{- \sqrt{2} b \phi} + (c^2 -3) | 1 - a e^{-b \phi/\sqrt{2}}|^2$ that resembles the Starobinsky potential~\cite{Starobinsky:1980te}.

\subsection{Polynomial potentials}

Though hybrid inflation needs several fields for its realisation, only the inflaton is dynamically relevant during inflation. Therefore, it makes sense to find generalisations of hybrid inflation in multi-field inflation by constructing new models of inflation where two or more fields are dynamical. As a simple case, let us consider a supergravity model with two superfields having a renormalisable (Wess-Zumino) superpotential, and
\begin{align}
    \label{eq:polynomial-K}
    K =& - \frac{1}{2} \sum_{i=1}^2 (\Phi^i - \bar{\Phi}^{\bar{i}}) + \cdots , \\
    \widetilde{W}=& \widetilde{W}_0 + \sum_i \mu_i \Phi^i + \frac{1}{2}\sum_{i,j} m_{ij}^2\Phi^i \Phi^j  + \frac{1}{3}\sum_{i,j,k} y_{ijk}\Phi^i \Phi^j \Phi^k,
\end{align}
where we have neglected the possible shift-symmetric kinetic mixing term $(\Phi^1-\bar{\Phi}^{\bar{1}})(\Phi^2 - \bar{\Phi}^{\bar{2}})$ without loss of generality.\footnote{Given the additional term $- \epsilon (\Phi^1-\bar{\Phi}^{\bar{1}})(\Phi^2 - \bar{\Phi}^{\bar{2}})$ with $\epsilon < 1$ in $K$, 
we define $\hat{\Phi}^1 \equiv \Phi^1 + \epsilon \Phi^2$ and $\hat{\Phi}^2 \equiv \sqrt{1 - \epsilon^2} \Phi^2$ 
so that the quadratic terms are $K = - (1/2) \sum_{i=1}^2 (\hat{\Phi}^i - \bar{\hat{\Phi}}^{\bar{i}})^2$. After dropping the hats, it reduces to Eq.~\eqref{eq:polynomial-K} but affects the quartic terms in $K$.} Each parameter is complex, while the overall constant phase of $W$ is unphysical. Two parameters can be removed by a constant real shift of $\Phi^i$ as $\Phi^i \to \Phi^i + a^i$ with $a^i \in \mathbb{R}$.  Therefore, there are $2 \times 10 - 1 - 2 = 17$ real free parameters.  

As a simple toy model, let us consider the minimal shift-symmetric $K$ (up to the quartic stabilisation terms) with the following superpotential:
\begin{align}
    \widetilde{W}=\frac{1}{2} m_\Phi \Phi^2 + \frac{1}{2}m_\Psi \Psi^2.
\end{align}
After the stabilisation, we have $\Phi^i - \bar{\Phi}^{\bar{i}}\simeq 0$, and the scalar potential is
\begin{align}
    V & = |m_\Phi|^2 |\Phi|^2 + |m_\Psi|^2 |\Psi|^2 + \frac{c_\Phi}{2} \text{Im}(m_\Phi \overline{m_\Psi} \Phi \bar{\Psi}^2) + \frac{c_\Psi}{2} \text{Im}(\overline{m_\Phi}m_\Psi \bar{\Phi}^2 \Psi) \nonumber \\
    & \quad + \frac{1}{4}(c_\Phi^2 + c_\Psi^2 -3)\left|m_\Phi\Phi^2 + m_\Psi \Psi^2\right|^2.
\end{align}
The cubic terms vanish if we assume $m_\Phi \overline{m_\Psi}$ to be real. Next, assuming $c_\Phi^2 + c_\Psi^2 -3 \simeq 0$, we can neglect the quartic term as well. Then we reproduce 
the simple two-field power-law potential again: 
\begin{align}
    V \simeq |m_\Phi|^2 |\Phi|^2 + |m_\Psi|^2 |\Psi|^2 
    \simeq \frac{1}{2} \left(|m_\Phi|^2 \phi^2 + |m_\Psi|^2 \psi^2 \right),
\end{align}
where again $\phi = \sqrt{2}\text{Re}\, \Phi$ and $\psi = \sqrt{2}\text{Re}\,\Psi$.

It is straightforward to extend this construction to any (large)  number of superfields in analogy with $N$-flation~\cite{Dimopoulos:2005ac}.  Let us consider a toy model with
\begin{align}
    K =& - \frac{1}{2}\sum_{i=1}^N (\Phi^i - \bar{\Phi}^{\bar{i}})^2 + \cdots , \\
    W =& e^{i \frac{c}{\sqrt{N}}\sum_{i=1}^N \Phi^i} \times \frac{m}{2} \sum_{i=1}^N (\Phi^i)^2,
\end{align}
where we have set $c_i = c/\sqrt{N}$ for simplicity and have introduced the universal mass parameter $m$ for $N$ shift-symmetric superfields $\Phi^i$. Each $c_i$ explicitly scales like $1/\sqrt{N}$, so that they are small in the large $N$ limit
when the number of superfields is $N\gg1$. In this case, the scalar potential reads
\begin{align}
    V & = e^K e^{i\frac{c}{\sqrt{N}} \sum_i (\Phi^i - \bar{\Phi}^{\bar{i}})} m^2 |\Phi^i|^2 \left[ 1 - i \frac{c}{2\sqrt{N}} \sum_j (\Phi^j - \bar{\Phi}^{\bar{j}}) 
 + \frac{c^2 - 3}{4} |\Phi^k|^2 \right] \nonumber \\
& = m^2 |\Phi^i|^2 \left(1 + \frac{c^2 -3}{4}|\Phi^k|^2 \right),
\end{align}
where the indices $i$ and $k$ are implicitly summed, and we have used $\Phi^i - \bar{\Phi}^{\bar{i}} = 0$ in the second equality. Without loss of generality, we can take  $K = 0$ after setting $\Phi^i - \bar{\Phi}^{\bar{i}} = 0$.
Therefore, if and only if 
$c^2 - 3 \geq 0$, the potential is kept non-negative in the large-field region. 

It is worth mentioning that single-large-field (high-scale) inflation usually implies inflaton field values of the order of $M_{\rm Pl}$, which makes it sensitive to quantum gravity constraints. In the Swampland Program~\cite{Vafa:2005ui, Ooguri:2006in}, the so-called distance conjecture~\cite{Ooguri:2006in} bounds from above the change of inflaton field value during inflation by approximately $10 M_{\rm Pl}$~\cite{Scalisi:2018eaz, Ketov:2025nkr, Furuta:2025tjt}. $N$-field inflation could be a solution to this obstruction by effectively dividing the inflaton change value by the square root of large $N$.

\subsection{Axionic models}

\subsubsection{Model with alignment mechanism}

\begin{figure}[tbhp]
    \centering
    \includegraphics[width=0.45\linewidth]{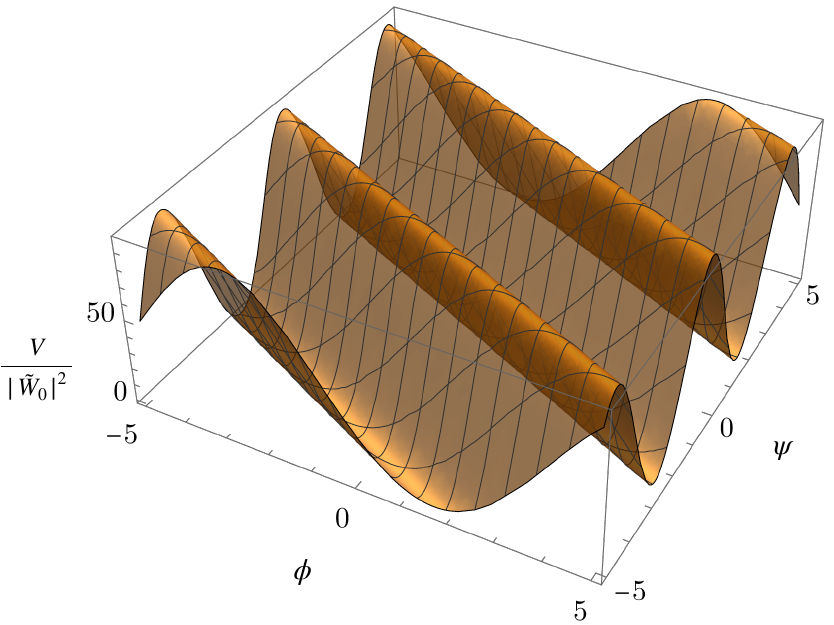}~
    \includegraphics[width=0.45\linewidth]{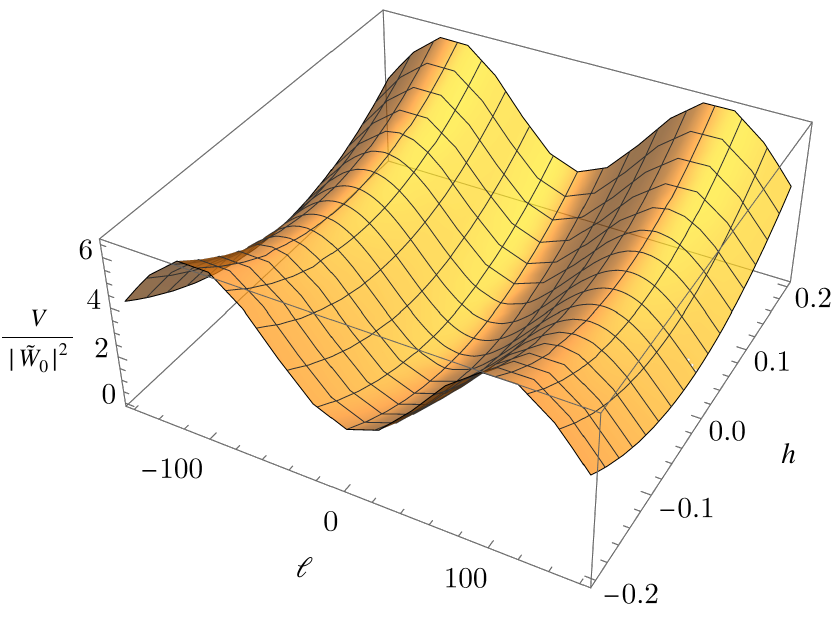}
    \caption{The normalized scalar potential~\eqref{V_KNP} with the KNP alignment mechanism. The left panel shows the potential in terms of $\phi= \sqrt{2} \text{Re}\,\Phi$ and $\psi =  \sqrt{2} \text{Re}\,\Psi$.  The right panel is in terms of the heavy component $h$ and the light component $\ell$ after the numerical diagonalisation to emphasise the potential along the light direction. Note that the scales are different in the $h$ and $\ell$ axes.  The parameters are $c = 2$, $\beta_\Phi =1$, $\beta_\Psi = 2$, $\beta'_\Phi = 2$, and $\beta'_\Psi = 4.1$. }
    \label{fig:V_KNP}
\end{figure}

Let us consider the following superpotential:
\begin{align}
    W = e^{i (c_\Phi \Phi + c_\Psi \Psi)} \widetilde{W}_0 \left[ 1 - \alpha e^{i(\beta_\Phi \Phi + \beta_\Psi \Psi)} - \alpha' e^{i(\beta'_\Phi \Phi + \beta'_\Psi \Psi)} \right],
\end{align}
where all $\beta$'s are real. It may look similar to the exponential superpotential~\eqref{W_plateau}, but here the exponents include the imaginary unit $i$, so it develops a sinusoidal or axionic scalar potential in the real directions.  We further assume that $(\beta_\Phi \Phi + \beta_\Psi \Psi)$ and $(\beta'_\Phi \Phi + \beta'_\Psi \Psi)$ are approximately linearly dependent to increase the effective decay constant. 
As in the previous example, we impose $\widetilde{W}=\widetilde{W}_\Phi = 0$ at $\Phi = \Psi = 0$ to restore supersymmetry after inflation. Then, because of the approximate linear dependence, $\widetilde{W}_\Psi$ approximately vanishes too. Then we find $\alpha$ and $\alpha'$ as
\begin{align}
    \alpha =& \frac{\beta'_\Phi}{\beta'_\Phi - \beta_\Phi}  \simeq \frac{\beta'_\Psi}{\beta'_\Psi - \beta_\Psi} ~, \\
    \alpha' =& 1 - \alpha = \frac{\beta_\Phi}{\beta_\Phi - \beta'_\Phi} \simeq \frac{\beta_\Psi}{\beta_\Psi - \beta'_\Psi} ~,
\end{align}
which are both real. Using real $\alpha$'s and $\beta$'s, and stabilising the imaginary parts of $\Phi$ and $\Psi$, we get the potential 
\begin{align}
    \frac{V}{|\widetilde{W}_0|^2}  =& 
c^2 - 3 + \alpha^2 (c^2 - 3 + \beta_\Phi^2 + \beta_\Psi^2) + \alpha'{}^2(c^2 -3 + \beta'_\Phi{}^2 + \beta'_\Psi{}^2) \nonumber \\
    & -2 \alpha (c^2 -3) \cos(\beta_\Phi \Phi + \beta_\Psi \Psi) - 2 \alpha' (c^2 -3) \cos (\beta'_\Phi \Phi + \beta'_\Psi \Psi) \nonumber \\
    & + 2 \alpha \alpha' (c^2 -3 + \beta_\Phi \beta'_\Phi + \beta_\Psi \beta'_\Psi) \cos \big[ (\beta'_\Phi - \beta_\Phi) \Phi + (\beta'_\Psi -\beta_\Psi)\Psi \big] . \label{V_KNP}
\end{align}
It follows that along the direction $\beta_\Psi \phi - \beta_\Phi \psi$ with $\phi = \sqrt{2}\text{Re}\,\Phi$ and $\psi = \sqrt{2}\text{Re}\,\Psi$, 
the field is approximately massless and the effective axion decay constant is large, so it is capable of playing the role of the inflaton. 
This is known in the literature as the Kim-Nilles-Peloso (KNP) alignment mechanism~\cite{Kim:2004rp}. The potential is depicted in Fig.~\ref{fig:V_KNP}.

\subsubsection{\texorpdfstring{$R$}{R}-axion models}
\label{sssec:R-axion}

The exponential factor in the superpotential and the shift-symmetric K\"ahler potential suggest an interpretation of the axion-like particle as an $R$-axion related to the $R$-symmetry in supersymmetry.

The superpotential $\widetilde{W}$ is independent of the $R$-axion if the $R$-symmetry is exact. Then the $R$-axion supermultiplet will be responsible for satisfying the condition $c^2 - 3 \geq 0$, but the inflaton will be another field. For instance, let us consider the following 
K\"ahler potential and superpotential:
\begin{align}
    K =& - \frac{1}{2} (\Phi - \bar{\Phi})^2 + \widetilde{K}(\Psi - \bar{\Psi}) + \cdots, \\
    W =& e^{i c \Phi} \widetilde{W}(\Psi),
\end{align}
where the real (imaginary) part of $\Phi$ is the $R$-axion ($R$-saxion) and the real (imaginary) part of $\Psi$ is the inflaton (sinflaton). 
The scalar potential is given by
\begin{align}
    V = e^K e^{i c (\Phi-\bar{\Phi})} \left[ (|K_\Phi + i c|^2 - 3)|\widetilde{W}|^2 + K^{\bar{\Psi}\Psi }|\widetilde{W}_\Psi + K_\Psi \widetilde{W}|^2 \right].
\end{align}
Note that the imaginary ($R$-saxion) direction can be stabilised by the term $- (\Phi -\bar{\Phi})^4$ in $K$ with a non-vanishing  $W \neq 0$ when $\Phi$ breaks supersymmetry, or by the term $- (\Phi- \bar{\Phi})^2(\Psi- \bar{\Psi})^2$ in $K$ when $\Psi$ breaks supersymmetry. It is straightforward to extend this class of models by replacing $\Psi$ with multiple superfields $\Psi^i$. 

Constraints on $R$-axion are briefly summarised in Appendix~\ref{sec:R-axion_constraints}.

\subsection{More general K\"ahler potentials}

Though our considerations above are limited to the multi-field polynomial-type K\"ahler potentials, their more general forms can be introduced as well. For instance, the Starobinsky- and no-scale supergravity-inspired logarithmic K\"ahler potentials can be introduced in the form 
\begin{align}
    K =  - \sum_i 3 \alpha_i \log \left[ 1 - \frac{i}{\sqrt{3\alpha_i}} (\Phi^i - \bar{\Phi}^{\bar{i}} ) + \xi_i (\Phi^i - \bar{\Phi}^{\bar i})^4 + \cdots \right], 
\end{align}
with the real parameters $\alpha_i$ and $\xi_i$.\footnote{
The parameters $\alpha_i$ in front of the logarithm are inspired by the $\alpha$-attractor models~\cite{Kallosh:2013hoa, Kallosh:2013yoa, Galante:2014ifa, Carrasco:2015pla}, but there is an important difference. In the $\alpha$-attractor-type or typical no-scale-type models, the inflaton is usually identified with a field combination appearing in the argument of the logarithm.  Here, we identify the orthogonal direction in the field space with the inflaton, which is protected by the manifest shift symmetry. Our $\alpha_i$'s are similar to the parameters $c_i$ introduced above, as can be seen after Taylor expansion of the logarithm. In fact, the expressions below (after stabilisation of the imaginary parts) depend only on the combination $\alpha_i + c_i$, where $c_i$'s are the parameters in the superpotential $W=e^{i c_i \Phi^i}\widetilde{W}$.
} 

The kinetic term after stabilisation is approximately canonical,
\begin{align}
    \frac{\mathcal{L}_\text{kin}}{\sqrt{-g}} = - \delta_{i \bar{j}} \partial^\mu \bar\Phi^{\bar j} \partial_\mu \Phi^i.
\end{align}
In our framework, the K\"ahler potential becomes approximately constant during inflation due to the quartic stabilisation term respecting the shift symmetry.  
The scalar potential after stabilisation takes the form
\begin{align}
    V = \delta^{\bar{j} i} \left( \widetilde{W}_i \overline{\widetilde{W}}_{\bar{j}} + i \tilde{c}_i \widetilde{W} \overline{\widetilde{W}}_{\bar{j}} - i \tilde{c}_{\bar{j}} \overline{\widetilde{W}}\widetilde{W}_i + \tilde{c}_i \tilde{c}_{\bar{j}} |\widetilde{W}|^2 \right) - 3 |\widetilde{W}|^2,
\end{align}
where now $\tilde{c}_i = c_i - i K_i = c_i + \alpha_i$, and similarly $\tilde{c}_{\bar{j}} = c_j + \alpha_j$. Therefore, it is essentially the same as that in the polynomial setup. 
When using the real superpotential and requiring $(\tilde{c}_i )^2 = 3$, the potential is significantly simplified to the form
arising in global (or rigid) supersymmetry,
\begin{align}
    V = \sum_i \left| \widetilde{W}_i \right|^2 .
\end{align}
This provides a multi-superfield generalisation of the results in Ref.~\cite{Ketov:2014hya}. The logarithmic form of $K$ is not essential for that. A detailed structure of $K$ becomes more important when one considers interactions between the inflaton sector and other matter sectors.

\section{Stabilisation mechanism for multi-superfield inflation}
\label{sec:stabilization}

In the preceding Sections, we assumed the stabilisation works by relying on the known results in the single-superfield case. 
In this section, we study the role of the quartic terms in more detail.

We always assume a shift-symmetric $K = K(\Phi^i - \bar{\Phi}^{\bar{i}})$. As in the single-superfield case, we neglect the cubic terms in $K$.\footnote{
Effects of the residual cubic term as well as the higher-dimensional terms were studied in Ref.~\cite{Terada:2014gnl} for the single-superfield case. 
} We first briefly explain the underlying reason for this.  
The general shift-symmetric $K$, after canonically normalising the quadratic term for simplicity of the analyses, can be written as 
\begin{align}
    K = \hat K_0 + i \hat c_i \hat{I}^i - \frac{1}{2} \delta_{ij} \hat{I}^i \hat{I}^j + i \frac{\hat\tau_{ijk}}{6}\hat{I}^i \hat{I}^j \hat{I}^k - \frac{\hat \xi_{ijkl}}{12}\hat{I}^i \hat{I}^j \hat{I}^k \hat{I}^l + \cdots,
\end{align}
where the hat notation is introduced by anticipating the later field redefinition, and $\hat{I}^i \equiv \hat {\Phi}^i - \bar{\Phi}^{\bar{i}}$ is just for compact notation. Note that $\hat I^i$ is not a holomorphic field, so there is no distinction between holomorphic ($i$, $j$, $k$, \dots) and anti-holomorphic ($\bar{i}$, $\bar{j}$, $\bar{k}$, \dots) indices for the coefficients $\hat c_i$, $\delta_{ij}$, $\hat \tau_{ijk}$, and $\hat \xi_{ijkl}$ because they are contracted with $(\Phi^i - \bar{\Phi}^{\bar{i}})$. The indices of these coefficients are totally symmetric.  For sufficiently large $\xi$'s, the dominant part of the potential around the minimum reads
\begin{align}
    V \sim & e^K K^{i\bar{j}} D_i W \overline{D}_{\bar{j}}\overline{W} + \cdots \nonumber \\
    \sim & 3 \left(1 - 2 (\hat{c}_i + \hat{\tau}_{i\mathcal{G}\mathcal{G}})(\mathrm{Im}\, \Phi^i) + 4 \hat{\xi}_{ij \mathcal{G}\mathcal{G}} (\mathrm{Im}\,\Phi^i)(\mathrm{Im}\,\Phi^j) + \cdots \right) (H^2 + m_{3/2}^2) + \cdots,
\end{align}
where the subscript $\mathcal{G}$ denotes the (time-dependent) goldstino component, \textit{e.g.}, $\tau_{i\mathcal{G}\mathcal{G}}\equiv \tau_{ijk}f^j f^k$ and $\xi_{ij\mathcal{G}\mathcal{G}}\equiv \xi_{ijkl}f^k f^l$, with $f^i \equiv \langle G^i/\sqrt{G_j G^j} \rangle$ being the goldstino direction, and we used the Friedmann equation. As can be seen from the above equation, if $\hat{\tau}_{ijk}$'s are too large, the minimum of the imaginary parts is significantly deviated from the origin. Therefore, expanding $K$ around the minimum, the cubic terms should be negligible.\footnote{Precisely speaking, the linear term $\hat \kappa_i$ also shifts the minimum, which is more carefully studied in Subsec.~\ref{ssec:deviations}}
Thus, we assume that there is a point in the field space where the cubic terms are negligible. Around the point, or in terms of the shifted field $I^i \simeq \hat{I}^i - (\hat{\xi}^{-1})^{ij\mathcal{G}\mathcal{G}}\hat{\tau}_{j\mathcal{G}\mathcal{G}}/2\sqrt{2}$ where $\hat{\xi}^{-1}$ is the inverse matrix of $\hat{\xi}$ focusing on the $(i,j)$ indices, the cubic terms are approximately absent. Explicitly performing this shift will be difficult unless one knows all the higher-dimensional terms. The residual constant ($K_0$) and linear ($ic_i I^i$) terms can be transferred to the superpotential by the super-Weyl-K\"ahler transformation.  After canonical normalisation of the quadratic term, $K$ becomes of the form
\begin{align}
    K = - \frac{1}{2}\delta_{ij}I^i I^j - \frac{\xi_{ijkl}}{12} I^i I^j I^k I^l,
\end{align}
up to higher-dimensional operators, where $I^i \equiv \Phi^i - \bar{\Phi}^{\bar{i}}$.  
To the end of this Section, we use this form of $K$.

Since the curvature of the K\"ahler field space of the scalar non-linear sigma model in supergravity plays the important role here, it is worth recalling the basics of K\"ahler geometry. The K\"ahler metric $K_{i \bar{j}} = \partial^2 K/\partial\Phi^i \partial\bar\Phi^{\bar{j}}$ is given by the second derivative of a K\"ahler potential $K$. 
In our case, it is 
\begin{align}
    K_{i \bar{j}} = & \delta_{i \bar{j}} +  \xi_{ijkl}I^k I^l~.
\end{align}
Up to the first order in $\xi_{ijkl} I^k I^l$, the inverse matrix is 
\begin{align}
    K^{\bar{j}i} \simeq \delta^{\bar{j}i} -  \xi_{ijkl} I^k I^l.
\end{align}
The non-vanishing K\"ahler connection is $\Gamma^{i}_{jk} = K^{i\bar{l}}K_{\bar{l}jk}$. 
The K\"ahler connection up to the leading order reads  
\begin{align}
    \Gamma^{i}_{j k} \simeq &  \delta^{i\bar{l}}K_{\bar{l}jk} 
    = 2 \xi_{ijkl}I^l~.
\end{align}
The mixed components of holomorphic and anti-holomorphic indices of the K\"ahler connection vanish. The Riemann curvature of the K\"ahler manifold is given by $\mathcal R_{i\bar{k}j\bar{l}} = K_{i\bar{k}j\bar{l}} - K^{\bar{n}m}K_{m\bar{l}\bar{k}} K_{\bar{n}ij}$. 
Up to the leading order, it is given by
\begin{align}
    \mathcal R_{i\bar{k}j\bar{l}} \simeq & K_{i\bar{k}j\bar{l}} 
    = -2 \xi_{ijkl}.
\end{align}

The holomorphic sectional curvature along a direction $X^i$ is defined by
\begin{align}
    \mathcal{K}(X)= - \frac{\mathcal R_{i\bar{j}k\bar{l}}X^i \bar{X}^{\bar{j}}X^k \bar{X}^{\bar{l}}}
    {(K_{i\bar{j}}X^i \bar{X}^{\bar{j}})^2}~. 
\end{align}
In particular, the sectional curvature along the goldstino direction $f_i = \langle G_i/\sqrt{G_jG^j}\rangle$ is given by
\begin{align}
    \mathcal{K}(f) = - \frac{\mathcal R_{i\bar{j}k\bar{l}}f^i f^{\bar{j}}f^k f^{\bar{l}}}{(K_{i\bar{j}}f^i f^{\bar{j}})^2}~.
\end{align}
This result can be understood as the goldstino-direction component ($\mathcal{G} = f_i \Phi^i$) of $\xi_{ijkl}$, \textit{i.e.}, $\mathcal{K}(f) \simeq 2 \xi_{\mathcal{G}\mathcal{G}\mathcal{G}\mathcal{G}}= 2\xi_{ijkl}f^i f^j f^k f^l$ (see Appendix~\ref{sec:field_space_rotation} for details), 
whose sign depends upon a convention used in the literature.\footnote{
It is also instructive to note in the single-superfield case that the sign flips between the sectional curvature and the Ricci scalar in the field space as $\mathcal{R}=K^{\Phi\bar{\Phi}}K^{\Phi\bar{\Phi}}\mathcal{R}_{\Phi\bar{\Phi}\Phi\bar{\Phi}} = - \mathcal{K}(\Phi)$. } For instance, the sectional curvature for the $\alpha$-attractor 
\cite{Kallosh:2013hoa, Kallosh:2013yoa, Galante:2014ifa, Carrasco:2015pla} in our notation is $-2/(3\alpha)$, where $\alpha = 1$ corresponds to the no-scale models.  The K\"ahler manifolds of  the $\alpha$-attractor models have hyperbolic geometry, and the hyperbolic geometry has the negative sectional curvature in our notation. On the other hand, we should require $\mathcal{K}(f)$ to be positive for the stabilisation of the sinflaton.

Given $\xi_{\mathcal{G}\mathcal{G}\mathcal{G}\mathcal{G}} > 0$, the necessary condition from Ref.~\cite{Covi:2008cn} 
needed to avoid a tachyonic instability is obviously satisfied in this case because
\begin{align}
    \mathcal{K}(f) \simeq 2 \xi_{\mathcal{G}\mathcal{G}\mathcal{G}\mathcal{G}} \gtrsim - \frac{2}{3} \frac{1}{1 + \gamma} ~,
\end{align}
where $\gamma \equiv V/(3m_{3/2}^2 M_\text{Pl}^2) \simeq H^2/m_{3/2}^2$. 
More generally, the term 
\begin{align}
    \mathcal R_{i\bar{j}k \bar{l}}f^k f^{\bar{l}} \simeq & -2 \xi_{ijkl} f^k f^{\bar l}
    \label{RijGG}
\end{align}
contributes to the effective mass squared of $(i,\bar{j})$-component.  Since the shift symmetry is imposed on $K$, only the imaginary parts receive large contributions.  This can be checked by noticing that in the mass matrix component $V_{ij}$, 
there is a term
\begin{align}
    e^{G}G^k \nabla_i \nabla_j G_k =& - e^G G_{ijk\bar{l}}G^k G^{\bar{l}} + \cdots.
\end{align}
For comparison, $V_{i\bar{j}}$ contains the term $- e^G G_{i\bar{j}k\bar{l}}G^k G^{\bar{l}}$. Since $G_{ij k \bar{l}} = -G_{i\bar{j}k\bar{l}}$ in our setup, the $\xi$-dependent part of the mass matrix has the schematic form $\propto \begin{pmatrix} 1 & -1 \\ -1 & 1 \end{pmatrix}$. After its diagonalisation, the real part is massless at the given order, whereas the imaginary part is massive provided that $\xi_{ij\mathcal{G}\mathcal{G}}$ is positive.   More precisely, the $\xi$-dependent part, which is dominant for $\xi \gg 1$, of the mass squared matrix of the imaginary parts of the general $(i,j)$ component is
\begin{align}
   m^2_{ij} \simeq  12 (H^2 + m^2_{3/2}) \xi_{ij\mathcal{G}\mathcal{G}}~,
\end{align}
where $\xi_{i j\mathcal{G}\mathcal{G}} \equiv \xi_{i j k l }f^k f^{\bar{l}}$. 

Since the behaviour of a scalar perturbation in quasi-de Sitter spacetime is characterised by the index 
$\nu = \sqrt{9/4 - m^2 /H^2}$, the isocurvature perturbations are suppressed for 
$m^2 \gtrsim (3 H / 2)^2$.  Therefore,  the relevant constraints are given by
\begin{align}
    \xi_{ij\mathcal{G}\mathcal{G}} \gtrsim  \frac{3H^2}{16(H^2 + m_{3/2}^2)}. 
\end{align}
More precisely, the left-hand side should be interpreted as the relevant eigenvalue. 
Should these constraints be violated, the imaginary parts of the superfields also participate in multi-field dynamics.

%----------------------------------------------
\subsection{Calculation of field deviations}\label{ssec:deviations}

In this Subsection, we explicitly make a distinction between complex fields and their decomposition into real and imaginary components as $\Phi^i=(\phi^i + i \chi^i)/\sqrt{2}$, where both $\phi^i$ and $\chi^i$ are real. 
Let us evaluate the deviations of $\langle \chi^i\rangle$ from $0$. For this purpose, we expand the potential $V$ up to the second order with respect to $\chi$'s and find the critical points by taking  $V_{\chi^i} = 0$. 
The kinetic term up to the second order in fluctuations is
\begin{align}
    \mathcal{L}_\text{kin} =& - \frac{1}{2} \left(\delta_{ij} - 2 \xi_{ijkl}\chi^k \chi^l \right) (\partial^\mu \phi^i \partial_\mu \phi^j + \partial^\mu \chi^i \partial_\mu \chi^j).
\end{align}
We find the scalar potential as
\begin{align}
\label{eq:V-quadratic}
    V =& e^{\sum_k \left(- \sqrt{2} c_k \chi^k +  (\chi^k)^2 \right) } \left\{\left( \delta^{i\bar{j}} + 2 \xi_{ijkl}\chi^k \chi^l \right) \right. \nonumber \\
    & \left. 
    \times \left[\widetilde{W}_i + i(c_i - \sqrt{2}\chi^i) \widetilde{W} \right] \left[\overline{\widetilde{W}}_{\bar{j}} - i(c_j - \sqrt{2}\chi^j) \overline{\widetilde{W}} \right] - 3 |\widetilde{W}|^2 \right\}.
\end{align}
In addition to the explicit presence of $\chi$'s in the potential above, the functions like  $\widetilde{W}(\Phi)$ also depend on $\chi$. Let us focus on the explicit $\chi$-dependence.    When the derivative used to calculate $V_{\chi^i}=i (V_i - V_{\bar{i}})/\sqrt{2}$ hits the implicit dependence on $\chi$, those contributions add up to give a slow-roll-suppressed contribution unless the slow-roll-suppression originates from a significant cancellation between $V_i$ and $V_{\bar{i}}$ in $V_{\phi^i} = (V_i + V_{\bar{i}})/\sqrt{2}$. 

We decompose Eq.~\eqref{eq:V-quadratic} up to the second order in $\chi$'s as follows:
\begin{align}
    V = V^{(0)} + V^{(1)}_i\chi^i + \frac{1}{2} V_{ij}^{(2)}\chi^i\chi^j ,
\end{align}
where 
\begin{align}
    V^{(0)} =& \delta^{i\bar{j}}\left(\widetilde{W}_i + i c_i  \widetilde{W} \right)\left(\overline{\widetilde{W}}_{\bar{j}} - i c_j \overline{\widetilde{W}} \right) - 3 |\widetilde{W}|^2, \\
    V^{(1)}_i\chi^i = & -\sqrt{2}V^{(0)} c_k \chi^k - \sqrt{2}i \chi^i \widetilde{W} \left( \overline{\widetilde{W}}_{\bar{i}} - i c_i \overline{\widetilde{W}} \right) + \sqrt{2}i  \chi^i \overline{\widetilde{W}} \left( \widetilde{W}_{i} + i c_i \widetilde{W} \right), \\
    \frac{1}{2} V_{ij}^{(2)}\chi^i\chi^j =& V^{(0)} \left((c_k \chi^k)^2 + (\chi^l)^2 \right) - 2 (c_k\chi^k) i \chi^i   \left[\widetilde{W}\left( \overline{\widetilde{W}}_{\bar{i}} - ic_i \overline{\widetilde{W}} \right) -  \overline{\widetilde{W}} \left( \widetilde{W}_i + i c_i \widetilde{W} \right)\right] \nonumber \\
    & + 2\xi_{ijkl}\chi^k \chi^l \left(\widetilde{W}_i + ic_i \widetilde{W}  \right)\left(\overline{\widetilde{W}}_{\bar{j}} - i c_j \overline{\widetilde{W}}\right) + 2 (\chi^i)^2 |\widetilde{W}|^2.
\end{align}
Note that $V^{(0)}$, $V^{(1)}_i$ and $V^{(2)}_{ij}$ themselves depend on $\chi$'s. 
Up to the second order in $\chi$'s, the scalar potential is 
\begin{align}
    V= \left\langle V^{(0)}\right\rangle 
    + \left( \left\langle V^{(0)}_{,i}\right\rangle + \left\langle V^{(1)}_i\right\rangle \right)\chi^i 
    + \frac{1}{2} \left( \left\langle V^{(0)}_{,ij} \right\rangle + \left\langle V^{(1)}_{i,j} \right\rangle 
    + \left\langle V^{(1)}_{j,i} \right\rangle + \left\langle V^{(2)}_{ij} \right\rangle \right) \chi^i \chi^j,
\end{align}
where the angle brackets denote the expectation values on the slow-roll inflation background. 
Therefore, the stationary condition reads
\begin{align}
    0=V_{\chi^i} 
    =& \left( \left\langle V^{(0)}_{,ij} \right\rangle + \left\langle V^{(1)}_{i,j} \right\rangle 
    + \left\langle V^{(1)}_{j,i} \right\rangle + \left\langle V^{(2)}_{ij} \right\rangle \right) \chi^j 
    + \left\langle V^{(1)}_i \right\rangle , \label{chi_stationary_condition_0}
\end{align}
where the notation ${}_{,i}$ denotes the differentiation with respect to $\chi^i$. 
As mentioned above, $\left\langle V^{(0)}_{,i} \right\rangle$ is slow-roll-suppressed, so we neglect it.
Explicitly, the constant term reads
\begin{align}
    \left\langle V_i^{(1)} \right\rangle = -\sqrt{2}c_i \left(V^{(0)} + 2 |\widetilde{W}|^2 \right) - \sqrt{2}i \left (\widetilde{W}\overline{\widetilde{W}}_{\bar{i}}- \overline{\widetilde{W}} \widetilde{W}_i\right) ,
\end{align}
where $\chi$ should be set to $0$ in this expression.

The remaining task is to evaluate the coefficient of the linear term in Eq.~\eqref{chi_stationary_condition_0}. The first term $\left\langle V^{(0)}_{,ij} \right\rangle$ is the model-dependent mass term that is unrelated to $c$ or $\xi$. We assume there is no strong tachyonic instability in this term, $\left\langle V^{(0)}_{,ij} \right\rangle \gtrsim \mathcal{O}(H^2)$. Though there may be a large positive mass term that is helpful for stabilisation, we neglect it in order to derive a conservative upper bound on the deviation $|\langle \chi^i \rangle| \neq 0$. Then the first derivative of $V^{(1)}_i$ takes the form 
\begin{align}
    \left\langle V^{(1)}_{i,j} \right\rangle \simeq & \widetilde{W}_{j}\left(\overline{\widetilde{W}}_{\bar{i} } -ic_i \overline{\widetilde{W}}\right) - \widetilde{W}\left(\overline{\widetilde{W}}_{\bar{i}\bar{j}} - i c_i \overline{\widetilde{W}}_{\bar{j}}\right) + \overline{\widetilde{W}}_{\bar{j}}\left( \widetilde{W}_i + i c_i \widetilde{W}\right) - \overline{\widetilde{W}} \left( \widetilde{W}_{ij} + i c_i \widetilde{W}_j \right).
\end{align}
For instance, when looking at $i = j$ component and assuming reality of all the coefficients in the superpotential $\widetilde{W}$ for simplicity, the $c_i$-dependent terms cancel, and it gives $\left\langle V^{(1)}_{i,i} \right\rangle \simeq 2 |\widetilde{W}_i|^2 - \widetilde{W}\overline{\widetilde{W}}_{\bar{i}\bar{i}} - \overline{\widetilde{W}}\widetilde{W}_{ii}$. Though this contribution may not be small, it is at least $\xi$-independent, so it does not play an important role in the large $\xi$ limit. For simplicity, we neglect those terms too.  

Similarly, we take the $\xi$-dependent part in $V^{(2)}_{ij}$ and neglect the $\xi$-independent part. 
Then, the stationary condition leads to the following equation
\begin{align}
   12 \xi_{ij\mathcal{G} \mathcal{G}}(H^2 + m_{3/2}^2)\chi^{j} \approx \sqrt{2}c_{i}(3H^2 + 2 m_{3/2}^2) + \sqrt{2}i \left( \widetilde{W}\overline{\widetilde{W}}_{\bar{i}} - \overline{\widetilde{W}}\widetilde{W}_{i} \right),
\end{align}
where we used the Friedmann equation. 
Therefore, the imaginary part of $\chi^i$ is stabilised as 
\begin{align}
    \chi^i \approx  (\xi^{-1})_{ij\mathcal{G}\mathcal{G}} \frac{\sqrt{2}c_j (3 H^2 + 2 m_{3/2}^2) +\sqrt{2}i \left( \widetilde{W}\overline{\widetilde{W}}_{\bar{j}} - \overline{\widetilde{W}}\widetilde{W}_{j} \right)}{12  (H^2 + m_{3/2}^2)} 
    \approx \frac{  (\xi^{-1})_{ij\mathcal{G}\mathcal{G}} \, c_j }{2\sqrt{2} },
\end{align}
where $(\xi^{-1})_{ij\mathcal{G}\mathcal{G}}$ is the inverse matrix (having $(i,j)$ indices) of $\xi_{ij\mathcal{G}\mathcal{G}}$, and in the second approximate equality (it is optional), we further assume $H^2 \gg m_{3/2}^2$ with purely real $\widetilde{W}$.

Since there can be additional mass contributions for the non-sgoldstino components, it is more appropriate to write
\begin{align}
    |\chi^i| \lesssim \frac{(\xi^{-1}_{ij\mathcal{G}\mathcal{G}})[ |c_j| + \mathcal{O}(1)]}{2\sqrt{2}},
\end{align}
where positive eigenvalues of $\xi_{ij\mathcal{G}\mathcal{G}}$ (or its inverse) are needed for stabilisation.   We recall  that the sgoldstino direction is time-dependent and the inflaton direction may be slowly changing during multi-field inflation. 

In summary, we have just confirmed that the imaginary parts $\chi^i$, which are in the non-shift-symmetric directions in the scalar field space, can be stabilised by sufficiently large quartic terms.

%======================================================================
\section{Conclusion}
\label{sec:conclusion}

In this paper, we have extended our previous proposal for inflation in supergravity without stabiliser fields to a broader framework that allows for multiple superfields. Multiple superfields protected by approximate shift symmetry contain light fields that can participate in single- or multi-field inflationary dynamics.  We have illustrated the mechanism with various example models of inflation realised in our supergravity setup. We have also studied the stabilisation mechanism for the directions not protected by shift symmetry and estimated the suppressed values of the field deviations and the masses of the stabilised components. 

Let us emphasise again the significance of our results and locate them in the context of inflation in supergravity.  First, our proposal is not just specific \emph{models} but a \emph{mechanism} that can accommodate many inflation models. Second, there are only several known \emph{mechanisms} that embed generic inflation models into supergravity such as $D$-term inflation models~\cite{Binetruy:1996xj, Halyo:1996pp, Binetruy:2004hh, Domcke:2017xvu, Kadota:2007nc, Kawano:2007gg, Kadota:2008pm, Farakos:2013cqa, Ferrara:2013rsa, Ketov:2019toi} and models with stabiliser fields (original models mainly characterised by superpotential~\cite{Kawasaki:2000yn, Kawasaki:2000ws, Kallosh:2010ug, Kallosh:2010xz} and those by K\"ahler potential~\cite{McDonough:2016der, Kallosh:2017wnt} (see also their cousins~\cite{Cribiori:2017laj, Kuzenko:2018jlz, Antoniadis:2018cpq, Aldabergenov:2018nzd, Antoniadis:2019nwz, Kuzenko:2019vaw, Farakos:2018sgq, Jang:2021vpb})).  Thus, our proposal significantly broadens the model-building possibilities for inflation in supergravity. 

The existence of several ways of embedding a given non-supersymmetric inflation model into supergravity has both advantages and disadvantages.  Even pinning down a specific viable inflation model in supergravity is a hard task because of a limited number of observables and several uncertainties, whereas  distinguishing different realisations of inflation in supergravity is an even harder task.  Nevertheless, exploring more possibilities in supergravity is important.  Moreover, having several realisations of a given inflation scenario at the phenomenological supergravity level is beneficial as the stepping stones toward a more fundamental description of inflation in quantum gravity or superstring theory.

There are several directions for future work: 
\begin{itemize}
    \item One direction is to pursue other possibilities to realise multi-field inflation in supergravity.  For example, a possibility is to reduce the magnitude of the stabilisation parameters $\xi_{ijkl}$ so that the imaginary parts can also participate in the multi-field inflationary dynamics.  In such a case, larger values of $c_i$ will be required to compensate for the negative term $- 3 e^K |W|^2$  in the potential $V$. 
\item Another direction is to establish the low-energy effective field theory (EFT) of our framework, in which supersymmetry is non-linearly realised. The EFT for ultraviolet theories with multiple shift-symmetric superfields was conjectured in Ref.~\cite{Aoki:2021nna}. It will be interesting to study how the constraints depend, if at all, on the structure of $\xi_{ijkl}$. 
\item As yet another direction, one has to construct multi-field supergravity models that are consistent with the latest observations and have a superpotential that is either simple or consistent with superstring theory in our framework.  This task will require numerical investigations. Studying possible signals of isocurvature perturbations and non-Gaussianity is also important because such signals are characteristic of multi-field inflation.  
\end{itemize}

In conclusion, our proposed multi-superfield realisation of $F$-term inflation in supergravity without stabiliser fields opens up a new avenue in the inflationary model building in supergravity.

%======================================================================
\section*{Acknowledgements}

TT is deeply grateful to his late father, who passed away during this work, for his lifelong support.  
JG is supported in part by the Basic Science Research Program through the National Research Foundation of Korea under No.~RS-2024-00336507. 
SVK is partially supported by Tokyo Metropolitan University, the World Premier International Research Center Initiative, MEXT, Japan, and Tomsk State University under the development program Priority-2030. 
TT and  SVK are (partially) supported by the 34th Academic research grant FY 2024 (Natural Science) under No.~9284 from DAIKO FOUNDATION. 
JG is grateful to the Asia Pacific Center for Theoretical Physics for hospitality while this work was in progress.

\appendix

\renewcommand{\theequation}{\thesection.\arabic{equation}}

\section{Change of basis in the field space}
\label{sec:field_space_rotation}
\setcounter{equation}{0}

In this Appendix, we summarise the field space rotation and introduce some basis-independent quantities used in the main text.

Since we are interested in slow-roll inflation, we assume an inflationary or quasi-de Sitter background. Though the original definitions of the quantities like the goldstino direction and the holomorphic sectional curvature depend on fields, we may evaluate them on a slowly-changing or approximately static background. 

Supersymmetry is spontaneously broken during inflation. Since we deal with the $F$-term inflation models, let us also assume  there are no $D$-term contributions, or they are negligible, for simplicity. 

Following Ref.~\cite{Covi:2008cn}, let us  define 
\begin{align}
    f^i \equiv \frac{G^i}{\sqrt{G_jG^j}}~,
\end{align}
where $G \equiv K + \log |W|^2$,  the subscripts in $G_i = K_i + W_i/W$ denote the derivatives, and $G^i = K^{i \bar{j}}G_{\bar{j}}$. $G^i$ is related to the supersymmetry breaking $F$-term via $F^i = - e^{G/2}G^i$, so that the goldstino direction $f^i$ is proportional to the vector specified by the $F$-term supersymmetry breaking. The goldstino direction also depends on the supersymmetry breaking contribution from the kinetic term. However, since we follow the definitions of Ref.~\cite{Covi:2008cn} and assume slow-roll, let us neglect the latter contribution.  We are interested in the expectation value $\langle f^i\rangle$ evaluated on the inflationary background. 

Sgoldstino can be defined by the weighted sum of $\Phi^i$ along the direction of $f^i$, \textit{i.e.},
\begin{align}
    \mathcal{G}= \langle f_i \rangle \Phi^i . 
\end{align}
This can be viewed as part of a unitary rotation in field space,\footnote{\label{fn:U_holomorphy} We require the leading (unit matrix) part of the K\"ahler metric to be invariant under such transformation, \textit{i.e.}, $\delta_{i \bar{j}} = U^\dagger{}^{-1}_{\bar{j}}{}^{\bar{l}} \delta_{k \bar{l}} U^{-1}{}^k{}_i $. It follows $U^{-1}{}^i{}_j = \delta^{i \bar{k}}U^\dagger_{\bar{k}}{}^{\bar{l}}\delta_{j \bar{l}}$.} which is also holomorphic,~\footnote{The components of $U$ are to be regarded as complex numbers rather than fields because of the slow-roll assumption. Under the slow-roll assumption, we can also neglect the derivatives acting on the transformation matrix arising from the kinetic term. } i.e. $\widehat{\Phi}^i = U^i{}_j \Phi^j$ (this also implies $\bar{\widehat{\Phi}}^{\bar{i}} = \bar{\Phi}^{\bar{j}} U^\dagger{}_{\bar{j}}{}^{\bar{i}}$), with
\begin{align}
    U^i{}_j = &
    \begin{pmatrix}
         \delta^{i,1} \, f_j  \\  V^{i(>1)}{}_j 
    \end{pmatrix} ,
\end{align}
where $V$ is a $(N-1) \times N$ matrix whose row vectors are normalized and orthogonal to $f_i$, \textit{i.e.}, $f^j V^i_j = 0$ but otherwise arbitrary. In this basis, the sgoldstino is the first component $\widehat{\Phi}^1 = \mathcal{G}$, so that $|f^1| = 1$ whose phase depends on the phase of $G^1$, and $f^i = 0$ ($i\neq 1$) up to the corrections of the order $\mathcal{O}(\xi (\Phi-\bar{\Phi})^2)$.  The inverse transformation matrix is essentially given by $U^\dagger$, that is, 
\begin{align}
     U^{-1}{}^i{}_j = \begin{pmatrix}
        f^i \delta_{j,1}  & V^\dagger{}^i_j 
    \end{pmatrix},
\end{align}
where we have implicitly converted the anti-holomorphic indices to the holomorphic ones by (the leading part of) the K\"ahler metric (see Footnote~\ref{fn:U_holomorphy}). 
The chain rule leads to 
\begin{align}
    \frac{\partial}{\partial \mathcal{G}} =& \frac{\partial \Phi^i}{\partial \mathcal{G}} \frac{\partial}{\partial\Phi^i} + \frac{\partial \bar\Phi^i}{\partial \mathcal{G}} \frac{\partial}{\partial \bar \Phi^i} = f^i \frac{\partial}{\partial \Phi^i}~, \\
    \frac{\partial}{\partial \bar {\mathcal{G}}} =& \frac{\partial  \Phi^{i}}{\partial \bar{\mathcal{G}}} \frac{\partial}{\partial\Phi^i} + \frac{\partial \bar\Phi^i}{\partial \bar{\mathcal{G}}} \frac{\partial}{\partial \bar \Phi^i} = f^{\bar{i}} \frac{\partial}{\partial \bar \Phi^{\bar{i}}}~.
\end{align}

Using these relations, we rewrite the holomorphic sectional curvature along $f^i$ as follows:
\begin{align}
    \mathcal{K}(f^i) = - \frac{\mathcal{R}_{i\bar{j}k\bar{l}}f^i f^{\bar{j}} f^k f^{\bar{l}}}{(K_{i\bar{j}}f^i f^{\bar{j}})^2} 
    = - \frac{\mathcal R_{\mathcal{G}\bar{\mathcal{G}}\mathcal{G}\bar{\mathcal{G}}}}{(K_{\mathcal{G}\bar{\mathcal{G}}})^2}~~. \label{sectional_curvature_goldstino_appendix}
\end{align}
A way to rewrite this equation  can be described as follows.  After inserting $UU^\dagger = U^\dagger U = \mathbf{1}$ 
sandwich-wise, we can change the field basis as
\begin{align}
    \mathcal{K}(f^i) = - \frac{\mathcal R_{\hat i\hat {\bar{j}}\hat k\hat{\bar{l}}}f^{\hat i} f^{\hat{\bar{j}}} f^{\hat k} f^{\hat{\bar{l}}}}{(K_{\hat{i}\hat{\bar{j}}}f^{\hat i} f^{\hat{\bar{j}}})^2} 
    = - \frac{\mathcal R_{\hat 1\hat {\bar{1}}\hat 1\hat{\bar{1}}}f^{\hat 1} f^{\hat{\bar{1}}} f^{\hat 1} f^{\hat{\bar{1}}}}{(K_{\hat{1}\hat{\bar{1}}}f^{\hat 1} f^{\hat{\bar{1}}})^2}  
    = - \frac{\mathcal R_{\hat 1\hat {\bar{1}}\hat 1\hat{\bar{1}}}}{(K_{\hat{1}\hat{\bar{1}}})^2}~,  \label{sectional_curvature_rotated_basis}
\end{align}
where in the second equality, we have used the fact that $|f^{\hat{1}}|=1$ and $f^{\hat{0}}= 0$, and in the third equality, we have cancelled the factors $|f^{\hat 1}|^2$ in the numerator and the denominator.  Of course,  Eq.~\eqref{sectional_curvature_rotated_basis} is the same as Eq.~\eqref{sectional_curvature_goldstino_appendix}. 
The explicit forms of the hatted quantities in those equations are 
\begin{align}
    f^{\hat{i}} =& U^i{}_j f^j, \\
    f^{\hat{\bar{i}}} =& U^\dagger _{\bar{j}}{}^{\bar{i}} f^{\bar{j}}, \\
    K_{\hat i \hat{\bar{j}}} =& \delta_{i \bar{i}} \delta_{j\bar{j}} U^j{}_k \delta^{k\bar{k}}U^\dagger_{\bar{k}}{}^{\bar{j}} (= \delta_{ij}), \\
    \mathcal{R}_{\hat i\hat {\bar{j}}\hat k\hat{\bar{l}}} =&  \mathcal{R}_{m\bar{n}p\bar{q}} U^{-1}{}^m{}_i (U^{\dagger})^{-1}_{\bar{j}}{}^{\bar{n}} U^{-1}{}^p{}_k (U^{\dagger})^{-1}_{\bar{l}}{}^{\bar{q}} .
\end{align}
In particular, we find
\begin{align}
    \mathcal{R}_{\hat 1\hat {\bar{1}}\hat 1\hat{\bar{1}}} =& \mathcal{R}_{m \bar{n}p\bar{q}} f^m f^{\bar{n}} f^p f^{\bar{q}} (= - 2 \xi_{\hat{1}\hat{1}\hat{1}\hat{1}}) ,\\
    K_{\hat{1}\hat{\bar{1}}}=& K_{i\bar{j}}f^i f^{\bar{j}} (= 1).
\end{align}
We define $\xi_{\mathcal{G}\mathcal{G}\mathcal{G}\mathcal{G}}$ by the following basis-independent expression:
\begin{align}
    \xi_{\mathcal{G}\mathcal{G}\mathcal{G}\mathcal{G}} \equiv & \xi_{ijkl}f^i f^{\bar{j}}f^k f^{\bar{l}}.
\end{align}
Similarly, we use $\xi_{ij\mathcal{G}\mathcal{G}} \equiv \xi_{ijkl}f^k f^{\bar{l}}$ in the main text.

%-----------------------------

\section{Constraints on \texorpdfstring{$R$}{R}-axion}
\label{sec:R-axion_constraints}
\setcounter{equation}{0}

The $R$-axion models were introduced in Subsec.~\ref{sssec:R-axion}.  In the inflaton sector, we did not introduce explicit breaking of the $\mathrm{U}(1)_R$ symmetry. This symmetry may be explicitly broken either slightly in the inflaton sector or substantially in other sectors. An explicit breaking gives rise to the mass of the $R$-axion. 

When the $R$-axion $a = \sqrt{2}\text{Re}\,\Phi $ has substantial interactions with the visible sector, there are generic constraints on axion-like particles (ALPs)~\cite{AxionLimits}. However, when the $R$-axion is in a hidden sector, many of the ALP constraints can be evaded.  The relevant constraints are briefly described below. 
\begin{enumerate}
    \item As regards the Planck bound \cite{Planck:2018jri} on the isocurvature perturbations of cold dark matter, by using Ref.~\cite{Graham:2025iwx}, the standard constraint~\cite{Beltran:2006sq, Hertzberg:2008wr, Hamann:2009yf} reads
\begin{align}
H_\text{inf} < 2.8 \times 10^{-5} f_a \theta_\mathrm{i}~,    
\end{align}
where $H_\text{inf}$ is the Hubble scale during inflation, $f_a$ is the axion decay constant, and $\theta_\text{i}$ is the initial misalignment angle. The constraint becomes weaker when the decay constant becomes smaller after inflation~\cite{Graham:2025iwx}. 

 \item The dark matter overproduction constraint is relevant if the axion is stable.  To satisfy the constraint, the upper bound of the mass $m_a$ is given by~\cite{Workman:2022ynf}
    \begin{align}
        m_a \lesssim 4.7 \times 10^{-19} \mathrm{eV} \, \left( \frac{f_a}{10^{16}\, \mathrm{GeV}}\right)^{-4} \theta_\mathrm{i}^{-4} ,
    \end{align}
    where a possible time dependence of $m_a$ and anharmonicity were neglected. If relativistic axions are produced, they are also constrained by the dark radiation constraint. 
    
    \item The black hole superradiance constraint is applicable even for the hidden-sector ALPs. The constraint is relevant for the axion mass around $10^{-19}\, \mathrm{eV}$ and $10^{-12}\, \mathrm{eV}$~\cite{Mehta:2020kwu, Baryakhtar:2020gao, Unal:2020jiy, Hoof:2024quk, Witte:2024drg}. 
    
    \item Overproduction of gravitinos from the $R$-axion decay will put another constraint. By construction, we need a sufficiently large $|c|$ to boost the positive contribution to the inflaton potential.  This means that supersymmetry is broken by the supermultiplet of $\Phi$ at least during and soon after inflation. If an $R$-axion decay into a pair of gravitinos is kinematically allowed, the partial decay rate is enhanced by $(m_a/m_{3/2})^2$ as~\cite{Endo:2006zj, Nakamura:2006uc, Kawasaki:2006gs, Asaka:2006bv, Dine:2006ii,  Kawasaki:2006hm, Endo:2007sz, Addazi:2017ulg}
    \begin{align}
        \Gamma(a \to \psi_{3/2}\psi_{3/2}) = \frac{|c|^2 m_a^5}{288 \pi m_{3/2}^2 M_\text{Pl}^2}~.
    \end{align}
    Strictly speaking, the condition $G^iG_i = 3$ (or, equivalently, $V=0$) was assumed to derive this formula, but it may be violated soon after inflation.  Also, well after inflation,  the superpotential terms other than the inflaton sector may become non-negligible, so that $G_\Phi = i c$ used above will be modified. A precise estimate of the gravitino abundance in our scenario (either in the $R$-axion cases or in general) is beyond the scope of this work.  The fate of gravitinos and the resulting constraints on the $R$-axion parameters depend on the gravitino mass, the mass spectrum of lighter particles, and the thermal history of the Universe.   
\end{enumerate}

\bibliographystyle{utphys}
\bibliography{ref}

\end{document}